%% file: interspeech_22.tex
\documentclass[a4paper]{article}

\usepackage{INTERSPEECH2022}
\usepackage{siunitx}
\usepackage[acronym]{glossaries}
\include{glossaries}
\usepackage{mathtools}
\usepackage{siunitx}
\usepackage{xcolor}
\usepackage{tikz,pgfplots}
\usepackage{tikzscale}
\usepgfplotslibrary{groupplots}
\usetikzlibrary{arrows, arrows.meta, decorations.markings}
\usepackage[capitalise]{cleveref}
\usepackage{url}
\usepackage{import}
\usepackage[skip=6pt]{caption}

\input{tikz_macros.tex}

\input{colors.tex}

\pgfplotsset{compat=1.17}

\newcommand{\pspeak}[1]{\bm{s}_{#1}}
\newcommand{\parray}[1]{\bm{a}_{#1}}
\newcommand{\oarray}[1]{\theta_{#1}}
\newcommand{\sro}[2]{\varepsilon_{#1}[#2]}

\newcommand{\sr}[2]{f_{#1}[#2]}
\newcommand{\micsig}[3]{y_{#1, #2}#3}

\newcommand{\totaldelay}[2]{\tau_{#1}[#2]}

\title{A Meeting Transcription System for an Ad-Hoc Acoustic Sensor Network}
\name{Tobias Gburrek,  Christoph Boeddeker, Thilo von Neumann, Tobias Cord-Landwehr, \\Joerg Schmalenstroeer, Reinhold Haeb-Umbach}
\address{
	Paderborn University, Germany
}
\email{\{gburrek,  boeddeker,  vonneumann, cord, schmalen, haeb\}@nt.upb.de}

\begin{document}
\setlength{\textfloatsep}{8pt}
\setlength{\intextsep}{8pt}
\setlength{\abovedisplayskip}{3pt}
\setlength{\abovedisplayshortskip}{2pt}
\setlength{\belowdisplayskip}{3pt}
\setlength{\belowdisplayshortskip}{2pt}

\maketitle
\begin{abstract}
We propose a system that transcribes the conversation of a typical meeting scenario that is captured by a set of initially unsynchronized microphone arrays at unknown positions. It consists of subsystems for signal synchronization, including both sampling rate and sampling time offset estimation, diarization based on speaker and microphone array position estimation, multi-channel speech enhancement, and automatic speech recognition.
With the estimated diarization information, a spatial mixture model is initialized that is used to estimate beamformer coefficients for source separation.
Simulations show that the speech recognition accuracy can be improved by synchronizing and combining multiple distributed microphone arrays compared to a single compact microphone array. Furthermore, the proposed informed initialization of the spatial mixture model  delivers a clear performance advantage over random initialization. 
\end{abstract}

\input{sections/intro3}

\input{sections/problem}

\input{sections/system}
\input{sections/experiments}
\input{sections/conclusions}

\section{Acknowledgements}
Computational resources were provided by the Paderborn Center for Parallel Computing.
Funded by the Deutsche Forschungsgemeinschaft (DFG, German Research Foundation) - Projects 282835863 and 448568305.

\bibliographystyle{IEEEtran}

\bibliography{interspeech_22}

\end{document}

%% file: glossaries.tex
\newacronym{DNN}{DNN}{deep neural network}
\newacronym{CDR}{CDR}{coherent-to-diffuse power ratio}
\newacronym{DoA}{DoA}{direction of arrival}
\newacronym[firstplural=time differences of arrival (TDoAs)]{TDoA}{TDoA}{time difference of arrival}
\newacronym{MSE}{MSE}{mean-squared error}
\newacronym{MLP}{MLP}{multilayer perceptron}
\newacronym{CRNN}{CRNN}{convolutional recurrent neural network}
\newacronym[plural=GPs,firstplural=Gaussian processes (GPs)]{GP}{GP}{Gaussian process}
\newacronym{GRU}{GRU}{gated recurrent unit}
\newacronym{RIR}{RIR}{room impulse response}
\newacronym{AWGN}{AWGN}{additive white Gaussian noise}
\newacronym{SNR}{SNR}{signal-to-noise ratio}
\newacronym{STFT}{STFT}{short-time Fourier transform}
\newacronym[plural=PSDs,firstplural=power spectral densities~(PSDs)]{PSD}{PSD}{power sprectral density}
\newacronym{AE}{AE}{absolute error}
\newacronym{MAE}{MAE}{mean-absolute error}
\newacronym{PE}{PE}{position error}
\newacronym{MPE}{MPE}{mean position error}
\newacronym{WASN}{ASN}{acoustic sensor network}
\newacronym{OoR}{OoR}{out-of-range}
\newacronym{ASR}{ASR}{automatic speech recognition}
\newacronym{TDNN}{TDNN}{time delay neural network}
\newacronym{CNN}{CNN}{convolutional neural network}
\newacronym{WLS}{WLS}{weighted least squares}
\newacronym{LS}{LS}{least squares}
\newacronym{RANSAC}{RANSAC}{random sample consensus}

\newacronym{GARDE}{GARDE}{\textbf{G}eometry c\textbf{A}libration f\textbf{R}om \textbf{D}istance \textbf{E}stimates}
\newacronym{MDS}{MDS}{Multi Dimensional Scaling}

\newacronym{CWLS}{CWLS}{constrained weighted least squares}
\newacronym{CRLB}{CRLB}{Cramer-Rao lower bound}
\newacronym{RMSE}{RMSE}{root mean square error}

\newacronym{CDF}{CDF}{Cumulative distribution function}
\newacronym{CDF2}{CDF}{Cumulative distribution function}

\newacronym{ABEL}{ABEL}{averaged burst error length}
\newacronym{ACD}{ACD}{average coherence drift}
\newacronym{ADC}{ADC}{analog-digital converter}
\newacronym{APT}{APT}{averaged processing time}
\newacronym{ASN}{ASN}{acoustic sensor network}
\newacronym{ASNs}{ASNs}{acoustic sensor networks}
\newacronym{ASRC}{ASRC}{arbitrary sampling rate conversion}
\newacronym{ATD}{ATD}{time drift}
\newacronym{ATS}{ATS}{accumulating time shift}
\newacronym{BI}{BI}{band-limited interpolation}
\newacronym{BSS}{BSS}{blind source separation}
\newacronym{CCF}{CCF}{cross-correlation function}
\newacronym{CCF-2}{CCF-2}{secondary \gls{CCF}}
\newacronym{CD}{CD}{coherence drift}
\newacronym{CFM}{CFM}{coherence function maximization}
\newacronym{CM}{CM}{correlation maximization}
\newacronym{CSD}{CSD}{cross-spectral density}
\newacronym{CSD-2}{CSD-2}{secondary \gls{CSD}}
\newacronym{CTC}{CTC}{continuous-time conversion}
\newacronym{DCML}{DCML}{\gls{ML} method for dynamic conditions}
\newacronym{DB}{DB}{Data base}
\newacronym{DD}{DD}{digital-to-digital}
\newacronym{DDC}{DDC}{digital-to-digital converter}
\newacronym{DSP}{DSP}{digital signal processing}
\newacronym{DTC}{DTC}{discrete-time conversion}
\newacronym{DXCP}{DXCP}{double-cross-correlation processor}
\newacronym{DXCPPhaT}{DXCP-PhaT}{\gls{DXCP} with phase transform}
\newacronym{ECS}{ECS}{energy correlation score}
\newacronym{FCF}{FCF}{filter correlation function}
\newacronym{FF}{FF}{free-field}
\newacronym{FFT}{FFT}{fast Fourier transform}
\newacronym{FFT-DXCP}{FFT-DXCP}{\gls{FFT} domain \gls{DXCP}}
\newacronym{FO}{FO}{frame-oriented}
\newacronym{GCC}{GCC}{generalized cross-correlation}
\newacronym{GCPSD}{GCPSD}{generalized cross power sprectral density}
\newacronym{GCCF}{GCCF}{generalized \gls{CCF}}
\newacronym{GCCPhaT}{GCC-PhaT}{generalized cross-correlation with phase transform}
\newacronym{SRPPhaT}{SRP-PhaT}{steered-response power phase transform}
\newacronym{GCSD}{GCSD}{generalized cross-spectral density}
\newacronym{GE}{GE}{Gilbert-Elliott}
\newacronym{IBI}{IBI}{iterative \gls{BI}}
\newacronym{ICA}{ICA}{independent component analysis}
\newacronym{IML}{IML}{iterative \gls{ML}}
\newacronym{IR}{IR}{impulse response}
\newacronym{IIR}{IIR}{infinite impulse response}
\newacronym{IFFT}{IFFT}{inverse \gls{FFT}}
\newacronym{ISTFT}{ISTFT}{inverse \gls{STFT}}
\newacronym{LCD}{LCD}{least-squares coherence drift}
\newacronym{LPD}{LPD}{linear-phase drift}
\newacronym{LTI}{LTI}{linear time-invariant}
\newacronym{LTV}{LTV}{linear time-variant}
\newacronym{LTVIR}{LTV-IR}{linear time-variant impulse response}
\newacronym{ML}{ML}{maximum likelihood}
\newacronym{MS}{MS}{multi-stage}
\newacronym{MSC}{MSC}{magnitude-squared coherence}
\newacronym{OML}{OML}{optimized \gls{ML}}
\newacronym{OR}{OR}{outlier removal}
\newacronym{PHAT}{PHAT}{phase transform}
\newacronym{PL}{PL}{packet loss}
\newacronym{PolyFar}{PolyFar}{polyphase-Farrow}
\newacronym{ppb}{ppb}{part per billion}
\newacronym{ppm}{ppm}{parts \ per \ million}

\newacronym{RBI}{RBI}{recursive band-limited interpolation}
\newacronym{RF}{RF}{realtime factor}
\newacronym{RTP}{RTP}{Real-time Transport Protocol}
\newacronym{RTCP}{RTCP}{Real-time Transport Control Protocol}

\newacronym{SCOT}{SCOT}{smoothed coherence transform}
\newacronym{SINR}{SINR}{signal-to-interpolation-noise ratio}
\newacronym{SO}{SO}{sample-oriented}
\newacronym{SPIB}{SPIB}{signal processing information base}
\newacronym{SPL}{SPL}{signal packet loss}
\newacronym{SRC}{SRC}{sampling rate conversion}
\newacronym{SRO}{SRO}{sampling rate offset}
\newacronym{STO}{STO}{sampling time offset}
\newacronym{TCP}{TCP}{Transmission Control Protocol}
\newacronym{TD}{TD}{time domain}
\newacronym{TDDXCP}{TD-DXCP}{time domain \gls{DXCP}}
\newacronym{TDD}{TDD}{time-delay difference}
\newacronym{TDOA}{TDoA}{time-difference of arrival}
\newacronym{UDP}{UDP}{User Datagram Protocol}
\newacronym{VAD}{VAD}{voice activity detection}
\newacronym{SAD}{SAD}{sound activity detection}
\newacronym{WACD}{WACD}{weighted average coherence drift}
\newacronym{DWACD}{DWACD}{dynamic weighted average coherence drift}
\newacronym{WG}{WG}{weighting}
\newacronym{WLCD}{WLCD}{weighted \gls{LCD}}
\newacronym{WASNs}{WASNs}{wireless acoustic sensor networks}
\newacronym{WCP}{WCP}{wideband correlation processor}
\newacronym{XC}{XC}{cross-correlation}
\newacronym{WER}{WER}{word error rate}
\newacronym{TXCO}{TXCO}{temperature compensated crystal oscillator}

\newacronym{OU}{OU}{Ornstein-Uhlenbeck}

\newacronym[firstplural=time differences of flight (TDoFs)]{TDOF}{TDoF}{time difference of flight}
\newacronym[firstplural=times of flight (ToFs)]{TOF}{ToF}{time of flight}

\newacronym{CSS}{CSS}{continuous speech separation}
\newacronym{cpWER}{cpWER}{concatenated minimum-permutation word error rate}
\newacronym{DER}{DER}{diarization error rate}

%% file: tikz_macros.tex
\usepackage{tikz}
\usepackage{standalone}
\usepackage{pgfplots}

\usetikzlibrary{positioning,chains,calc,shapes.geometric, shadows, shapes.misc, fit, arrows, shapes.callouts}
\usepgfplotslibrary{groupplots}
\usetikzlibrary{backgrounds}

%% file: colors.tex
\definecolor{yellow}{RGB}{255, 198, 0}
\definecolor{orange}{RGB}{255, 130, 0}
\definecolor{blue}{RGB}{0, 32, 91}
\definecolor{red}{RGB}{198, 53, 39}
\definecolor{magenta}{RGB}{138, 27, 97}
\definecolor{lightblue}{RGB}{0, 159, 223}
\definecolor{green}{RGB}{0, 155,119}
\definecolor{lightgreen}{RGB}{132, 189,0}
\definecolor{greenyellow}{RGB}{208, 223,0}

%% file: sections/intro3.tex

\section{Introduction}
\label{sec:intro}   

This contribution is concerned with the diarization and transcription of meetings, where the considered meeting scenarios are characterized by conversations between a small but unknown number of participants at fixed positions.
Of those, none, one or even two speakers may be active at a time. 


We here consider an ad-hoc \gls{WASN} setup, where multiple initially unsynchronized microphone arrays at unknown positions are used for signal capture, which is different from most studies on meeting diarization and recognition \cite{Raj2021}.   
A notable exception is the meeting transcription system presented in~\cite{Araki18}, that also utilizes initially unsychronized distributed microphones.
As mentioned there, the microphones' and the speakers' position are typically not known beforehand which complicates the usage of spatial information for speech enhancement. 
Another problem making the usage of spatial information more difficult arises from the required signal synchronization.
Typical signal synchronization systems, also the system presented in \cite{Araki18}, do not differentiate between the contribution of time shifts caused by differing \glspl{TOF} of a signal to the microphones and time shifts caused by differing recording start times to the \gls{TDoA} between two microphones.
Thus, it is not guaranteed that the estimated \glspl{TDoA} between the synchronized signals correctly reflect the \glspl{TDOF} between the speakers' and the microphones' positions and, therefore, carry spatial information. 
Due to these facts the authors of~\cite{Araki18} did not exploit any spatial information for their speech enhancement subsystem and opted for a fully blind speech enhancement.

Here, we build upon the idea of the signal synchronization system, we proposed in~\cite{Gburrek22}, and support the signal synchronization by \gls{TOF} information in form of estimates of the distances between the microphones and the speakers~\cite{gburrek_21_dist}.
This enables to obtain synchronized signals maintaining \glspl{TDoA} which correctly represent the microphones' and speakers' positions.
Furthermore, a geometry calibration~\cite{GeoJournal} is performed to infer both the microphones' and the speakers' positions from the microphone signals.
Both, the physically correct signal synchronization and the knowledge about the microphone and speaker positions, are subsequently used to perform a multi-speaker tracking based on \gls{SRPPhaT}~\cite{DiBiase01}.
Since the speakers' positions also provide their identities for the considered scenario assuming fixed speaker positions, the multi-speaker tracking  corresponds to a diarization which is also able to cope with overlapping speech.
The diarization estimate of who speaks when is then taken to initialize a spatial mixture model~\cite{Ito2016cACGMM}, whose outcome in turn is used to compute beamformer coefficients for source separation and signal enhancement. 
Finally, the enhanced signals are forwarded to the speech recognizer.

In simulations we show that the distributed nature of the \gls{WASN} leads to an improved source separation, diarization and meeting transcription compared to a system utilizing a single microphone or a single compact microphone array.
Particularly, the usage of spatial diarization information to initialize the spatial mixture model leads to an improved speech enhancement compared to an uninformed initialization that sets the initial values of the class posterior probabilities to draws from a Dirichlet distribution. 
Moreover, the results demonstrate that the proposed transcription system achieves nearly the same performance as a system that uses oracle speaker activity information and perfectly synchronous signals.

The remainder of the paper is structured as follows:
In \Cref{sec:problem} the considered meeting scenario is defined. 
Afterwards, the proposed meeting transcription system for ad-hoc \glspl{WASN} is introduced in \Cref{sec:system}.
An investigation of the proposed meeting transcription system is presented in \Cref{sec:exp}.
Finally, we end with the conclusions drawn in \Cref{sec:conclusions}.

%% file: sections/problem.tex
\section{Problem statement}
\label{sec:problem}

\begin{figure}[htb]
	\centering
	\def\svgwidth{.8\columnwidth}
	\small
	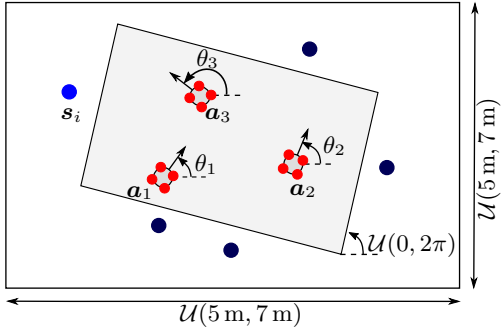
	\caption{Simulated recording setup including $J{=}3$ microphone arrays at positions $\parray{j}$ with  orientations $\oarray{j}, \ j \in \{1, 2, 3\}$, on a table. Figure not at scale; red dots: microphones; dark blue dots: speakers; blue dot: $i$-th speaker at position $\pspeak{i}$.}
	\label{fig:setup}
\end{figure}

We consider a meeting scenario as shown in Fig.~\ref{fig:setup} with $I$ speakers sitting at fixed but unknown positions $\pspeak{i}$, $i  \in  \{1, 2, \dots, I\}$, around a table in a reverberant room.
Although most of the time only one speaker is active, there are also quiet periods and times when up to two speakers are active at the same time.
The meeting is recorded by an ad-hoc \gls{WASN} formed by $J$ compact microphone arrays. 
All microphone arrays are placed on a table at  fixed but unknown positions $\parray{j}$ with an orientation $\oarray{j}$ , $j  \in  \{1, 2, \dots, J\}$.
Each microphone array consists of $M \geq 3$ microphones that do not lie on a line.
Moreover, we assume the arrangement of the microphones within an  array to be known.

Due to the independent hardware of the microphone arrays all  arrays start recording the meeting at a different point in time, causing a \gls{STO} $T_j$~\cite{Gburrek22}.
Furthermore, the frequencies of the clocks driving the sampling processes of the different microphone arrays will slightly differ from the nominal sampling frequency $f_s$ and also be time-varying such that all microphones of the $j$-th microphone array are sampled with a sampling rate $\sr{j}{n} {=} (1 {+} \sro{j}{n}) {\cdot} f_s$~\cite{Gburrek22}.
Here, $\sro{j}{n}$ denotes the time-varying \gls{SRO} of the \mbox{$j$-th} microphone array and $n$ the discrete-time sample index.


Sampling the continuous-time signal $\micsig{j}{m}{(t)}$ recorded by the $m$-th sensor of the $j$-th microphone array gives the following discrete-time signal~\cite{Gburrek22}:
\begin{align}
\micsig{j}{m}{[n]} = \micsig{j}{m}{\left(\frac{n}{f_s} {-} \frac{1}{f_s} {\cdot} \left(\smash[b]{\underbrace{{-}T_j \cdot f_s + \sum_{\tilde{n}=0}^{n-1} \sro{j}{\tilde{n}}}_{\coloneqq \totaldelay{j}{n}}}\right)\right)},
\vphantom{\underbrace{+ T_l - \sum_{\tilde{n}=0}^{n-1} \frac{\sro{i}{\tilde{n}}}{f_s}}_{\coloneqq - \totaldelay{j}{n} / f_s}}
\label{eq:async}
\end{align} 
i.e., the sampling time of the $n$-th sample is shifted by $\totaldelay{j}{n}$~\si{samples} w.r.t.~a signal sampled using a perfect clock ($T_j {=} \SI{0}{\second}$, $\sro{j}{n} {=} \SI{0}{ppm}$).

%% file: figures/scenario.pdf_tex
\begingroup%
  \makeatletter%
  \providecommand\color[2][]{%
    \errmessage{(Inkscape) Color is used for the text in Inkscape, but the package 'color.sty' is not loaded}%
    \renewcommand\color[2][]{}%
  }%
  \providecommand\transparent[1]{%
    \errmessage{(Inkscape) Transparency is used (non-zero) for the text in Inkscape, but the package 'transparent.sty' is not loaded}%
    \renewcommand\transparent[1]{}%
  }%
  \providecommand\rotatebox[2]{#2}%
  \newcommand*\fsize{\dimexpr\f@size pt\relax}%
  \newcommand*\lineheight[1]{\fontsize{\fsize}{#1\fsize}\selectfont}%
  \ifx\svgwidth\undefined%
    \setlength{\unitlength}{1032.54503145bp}%
    \ifx\svgscale\undefined%
      \relax%
    \else%
      \setlength{\unitlength}{\unitlength * \real{\svgscale}}%
    \fi%
  \else%
    \setlength{\unitlength}{\svgwidth}%
  \fi%
  \global\let\svgwidth\undefined%
  \global\let\svgscale\undefined%
  \makeatother%
  \begin{picture}(1,0.65506524)%
    \lineheight{1}%
    \setlength\tabcolsep{0pt}%
    \put(0,0){\includegraphics[width=\unitlength,page=1]{scenario.pdf}}%
    \put(0.24898263,0.25116727){\color[rgb]{0,0,0}\makebox(0,0)[lt]{\lineheight{1.25}\smash{\begin{tabular}[t]{l}$\bm{a}_1$\end{tabular}}}}%
    \put(0.58476662,0.25565654){\color[rgb]{0,0,0}\makebox(0,0)[lt]{\lineheight{1.25}\smash{\begin{tabular}[t]{l}$\bm{a}_2$\end{tabular}}}}%
    \put(0.9811737,0.55945356){\color[rgb]{0,0,0}\makebox(0,0)[lt]{\begin{minipage}{0.02905442\unitlength}\raggedright \end{minipage}}}%
    \put(0.41011364,0.41212782){\color[rgb]{0,0,0}\makebox(0,0)[lt]{\lineheight{1.25}\smash{\begin{tabular}[t]{l}$\bm{a}_3$\end{tabular}}}}%
    \put(0,0){\includegraphics[width=\unitlength,page=2]{scenario.pdf}}%
    \put(0.39293522,0.52295991){\color[rgb]{0,0,0}\makebox(0,0)[lt]{\lineheight{1.25}\smash{\begin{tabular}[t]{l}$\theta_3$\end{tabular}}}}%
    \put(0,0){\includegraphics[width=\unitlength,page=3]{scenario.pdf}}%
    \put(0.65353142,0.34271215){\color[rgb]{0,0,0}\makebox(0,0)[lt]{\lineheight{1.25}\smash{\begin{tabular}[t]{l}$\theta_2$\end{tabular}}}}%
    \put(0.38978238,0.30404994){\color[rgb]{0,0,0}\makebox(0,0)[lt]{\lineheight{1.25}\smash{\begin{tabular}[t]{l}$\theta_1$\end{tabular}}}}%
    \put(0.11396575,0.41196394){\color[rgb]{0,0,0}\makebox(0,0)[lt]{\lineheight{1.25}\smash{\begin{tabular}[t]{l}$\bm{s}_i$\end{tabular}}}}%
    \put(0,0){\includegraphics[width=\unitlength,page=4]{scenario.pdf}}%
    \put(0.34774185,-0.00505806){\color[rgb]{0,0,0}\makebox(0,0)[lt]{\lineheight{1.25}\smash{\begin{tabular}[t]{l}$~\mathcal{U}(\SI{5}{m},\SI{7}{m})$\end{tabular}}}}%
    \put(0,0){\includegraphics[width=\unitlength,page=5]{scenario.pdf}}%
    \put(1.00554921,0.19384515){\color[rgb]{0,0,0}\rotatebox{90}{\makebox(0,0)[lt]{\lineheight{1.25}\smash{\begin{tabular}[t]{l}$~\mathcal{U}(\SI{5}{m},\SI{7}{m})$\end{tabular}}}}}%
    \put(0,0){\includegraphics[width=\unitlength,page=6]{scenario.pdf}}%
    \put(0.73543975,0.13797602){\color[rgb]{0,0,0}\makebox(0,0)[lt]{\lineheight{1.25}\smash{\begin{tabular}[t]{l}$~\mathcal{U}(0,2\pi)$\end{tabular}}}}%
    \put(0,0){\includegraphics[width=\unitlength,page=7]{scenario.pdf}}%
  \end{picture}%
\endgroup%

%% file: sections/system.tex

\section{Meeting transcription system}
\label{sec:system}

\begin{figure}[t!]
	\centering
	\scriptsize
	\input{figures/system}
	\caption{Meeting transcription for ad-hoc \glspl{WASN}. Double arrows: audio signals; single arrows: estimated information}
	\label{fig:system}
\end{figure}
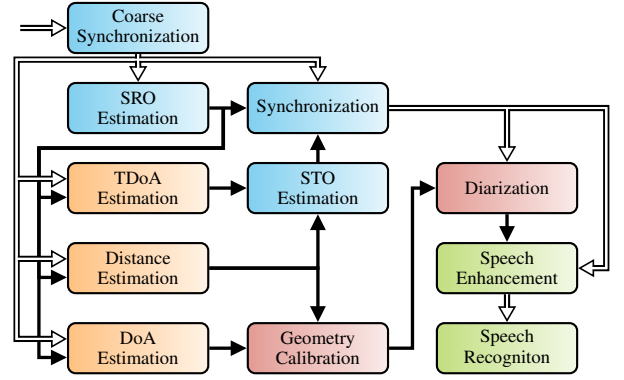

\Cref{fig:system} shows the block diagram of the proposed meeting transcription system.
As a first step, the signals recorded by different microphone arrays are synchronized (blue blocks in \cref{fig:system}).
Next, a diarization (red blocks in \cref{fig:system}) is performed based on speaker position information gathered from a multi-source localization.
The resulting speaker diary is utilized to initialize a multi-channel speech enhancement system whose output is fed to the~\gls{ASR} system (green blocks in \cref{fig:system}).

\subsection{Signal synchronization}
\label{subsec:sync}
Without loss of generality, all signals are synchronized w.r.t. the first microphone array. 
Moreover, we only use the first channel of the microphone arrays to estimate their \glspl{SRO} and \glspl{STO} w.r.t. the first microphone array.
First, the signals are coarsely synchronized based on the maximum of the cross-correlation between the first \SI{20}{\second} of the signals~\cite{Gburrek22}. 
This is to ensure that the $\ell$-th signal frames which are extracted from microphones belonging to different arrays in the following subsystems of the transcription system, roughly contain the same segment of the source signal.
Afterwards the \glspl{SRO} and \glspl{STO} between the microphone arrays are estimated and compensated.
We employ the \gls{DWACD} method, which we proposed in~\cite{Gburrek22}, for \gls{SRO} estimation.
To compensate for the \glspl{SRO} we utilize the \gls{STFT}-resampling method from~\cite{Schmalenstroeer2018}. 


\subsubsection{Sampling time offset}
\label{subsubsec:sto}
The \gls{TDOA}  between the signals recorded at two microphones from  different microphone arrays is the sum of two contributions:  the
 \gls{TDOF} caused by the differences in distance between the speaker and the microphones, and the \gls{STO} that reflects the different start times of the recording at  microphone arrays.
Here, we intend to use spatial information for diarization and therefore wish to compensate only for the \gls{STO}  and keep  the \glspl{TDOF} unmodified~\cite{Gburrek22}.
 After compensating for the \gls{SRO}, the \gls{TDOA} between the first channel of the $j$-th microphone array and the reference channel, i.e., the first channel of the first microphone, is given by
\begin{align}
	\tau_{1,j,i} = \frac{d_{j,1,i}- d_{1,1,i}}{c} - \left(T_j -T_1\right)  = \delta_{1,j,i} - T_{1,j},
	\label{eq:shift_tdof_sto}
\end{align}
if the $i$-th speaker is active~\cite{Gburrek22}.
Here, $c$ denotes the speed of sound and $d_{j,m,i}$ is the distance between the $i$-th speaker and the $m$-th microphone of the $j$-th microphone array. 
In~\eqref{eq:shift_tdof_sto}, the fraction corresponds to the \gls{TDOF} and the term in parentheses, $T_{1,j} {=} \left(T_j -T_1\right)$, to the \gls{STO}. 

The \gls{LS}-based \gls{STO} estimator from~\cite{Gburrek22} does not account for unbalanced activities of the speakers such that a speaker who speaks more has a larger influence on the \gls{STO} estimate.
Therefore, value pairs consisting of a frame-wise \gls{TDOF} estimate $\widehat{\delta}_{1,j}[\ell]$, which results from the distance estimates, and  the corresponding \gls{TDoA} estimate $\widehat{\tau}_{1,j}[\ell]$ for the same frame, are clustered to summarize frames belonging to the same speaker position.
To do so, first, the frame-wise pairs are clustered on the basis of the \gls{TDoA} estimates $\widehat{\tau}_{1,j}[\ell]$.
Subsequently, the frame-wise estimates within each time shift cluster are clustered on the basis of the \gls{TDOF} stimates $\widehat{\delta}_{1,j}[\ell]$.
Each tuple of \gls{TDOF} cluster and corresponding time shift cluster now represents an \gls{STO} candidate (see~\eqref{eq:shift_tdof_sto}).
Finally, these tuples are clustered on the basis of the associated \gls{STO} value.
The \gls{STO} estimate is given by the \gls{STO} value belonging to the cluster with the highest cardinality.

We employ the \gls{GCCPhaT}~\cite{Knapp1976} to estimate the \gls{TDOA}.
The distances between the speakers and the microphone arrays are estimated using the \acrlong{DNN} based estimator from~\cite{gburrek_21_dist}.
To compensate for the \gls{SRO}, the shift of the analysis window of both estimators is adapted in the same way as in the \gls{WACD} method (see~\cite{Gburrek22}).

\subsection{Spatial meeting diarization}
\label{subsec:srp}
Assuming fixed speaker positions, the positions of the active speakers  provide their identity. Meeting diarization can thus be performed by a speaker localization system. We employ \gls{SRPPhaT} because it is able to localize multiple simultaneously active acoustic sources. However, 
\gls{SRPPhaT} requires the knowledge of the relative position between the microphone arrays, which first needs to be estimated.
Thus, the first step is a geometry calibration, i.e., the estimation of the positions $\parray{j}$ and orientations $\oarray{j}$ of the microphone arrays, which we achieve using the iterative data set matching method described in ~\cite{GeoJournal}.
The iterative data set matching method takes \gls{DoA} estimates obtained using the complex Watson kernel method~\cite{drudecwk2015} and estimates of the distances between the speaker and the microphone arrays as input.

In addition to the geometry of the \gls{WASN}, the iterative data set matching method also provides robust estimates of the speakers' positions $\widehat{\bm{s}}_{i}$~\cite{GeoJournal}. Those are used to support the multi-source localization. 
On the one hand, the speaker position estimates $\widehat{\bm{s}}_{i}$ are utilized for a robust single-speaker tracking whose results are subsequently refined using \gls{SRPPhaT} to be able to cope with speaker overlap.
On the other hand, these estimates are used as a-priori knowledge for \gls{SRPPhaT} to limit the number of grid points searched for speaker activity.

For single speaker-tracking, the frame-wise speaker activity is estimated using an energy-based \gls{VAD}.
If speech is detected in the $\ell$-th frame, the \gls{DoA} and distance estimates for the  $\ell$-th frame are utilized together with the estimate of the geometry of the \gls{WASN} to obtain the speaker position estimate $\widehat{\bm{s}}[\ell]$ (median of the positions w.r.t. the coordinate system of the single microphone arrays, which are estimated from the \gls{DoA} and distance, after being mapped to the global coordinate system~\cite{GeoJournal}).
The speaker at position $\widehat{\bm{s}}_{i}$ is declared to be active for the $\ell$-th frame if $\widehat{\bm{s}}_{i}$ is the nearest speaker position estimate w.r.t. $\widehat{\bm{s}}[\ell]$ and if the distance between $\widehat{\bm{s}}[\ell]$  and $\widehat{\bm{s}}_{i}$ is below a given threshold. 
This results in a first frame-wise activity estimate $\widehat{a}_{i}[\ell]$ for each speaker.



In a second step \gls{SRPPhaT}, i.e., a multi-speaker tracking method, is used to add the activity of any additional active speakers in a time frame to the activity estimates $\widehat{a}_{i}[\ell]$.
\gls{SRPPhaT} is based on the calculation of the steered response power $P[\ell, \widehat{\bm{s}}_u^\text{SRP}]$~\cite{DiBiase01} for a set of $U$ speaker position candidates $\widehat{\bm{s}}_u^\text{SRP}$, $u \in \{1,2, \dots, U\}$, by accumulating the pair-wise \gls{GCCPhaT} values of all microphone pairs, where the \gls{GCCPhaT} functions are evaluated at the time lag corresponding to the theoretical \glspl{TDOF} belonging to the position $\widehat{\bm{s}}_u^\text{SRP}$.


Due to errors of the \gls{SRO} or \gls{STO} estimates, small time shifts might remain between the signals after synchronization in addition to the \glspl{TDOF}.
To account for these time shifts a grid of positions around each position estimate $\widehat{\bm{s}}_{i}$ is used rather than using a single position candidate for each speaker position estimate $\widehat{\bm{s}}_{i}$.
Afterwards, the steered response powers for each position within each grid belonging to the speaker position $\widehat{\bm{s}}_{i}$ are accumulated, leading to the power $P_\text{total}[\ell, \widehat{\bm{s}}_{i}]$.
A speaker at position $\widehat{\bm{s}}_{i}$ is declared to be active in time frame $\ell$ in addition to an already detected speaker if the power $P_\text{total}[\ell, \widehat{\bm{s}}_{i}]$ is larger than a given threshold. 
Finally, the activity estimates $\widehat{a}_{i}[\ell]$  are temporally smoothed.

\input{sections/source_sep}

%% file: figures/system.tex
\tikzset{%
  block/.style    = {draw, thick, rectangle, minimum height = 2.7em, minimum width = 7.5em, fill=white, align=center, rounded corners=0.1cm},
  sum/.style      = {draw, circle, node distance = 2cm}, 
  cross/.style={path picture={\draw[black](path picture bounding box.south east) -- (path picture bounding box.north west)
		 (path picture bounding box.south west) -- (path picture bounding box.north east);}},
	   zigzag/.style = {
	   	to path={ -- ($(\tikztostart)!.55!-9:(\tikztotarget)$) --
	   		($(\tikztostart)!.45!+9:(\tikztotarget)$) -- (\tikztotarget)
	   		\tikztonodes},sharp corners}
               }
\tikzstyle{branch}=[{circle,inner sep=0pt,minimum size=0.3em,fill=black}]
\tikzstyle{box} = [draw, dotted, inner xsep=4mm, inner ysep=3mm]
\tikzstyle{every path}=[line width=0.1em]
\tikzstyle{vecArrow} = [thick, decoration={markings,mark=at position
   1 with {\arrow[semithick]{open triangle 60}}},
   double distance=.75pt, shorten >= 5.5pt,
   preaction = {decorate},
   postaction = {draw,line width=.75pt, white,shorten >= 4.5pt}]
\tikzstyle{innerWhite} = [semithick, white,line width=.75pt, shorten >= 4.5pt]
\tikzstyle{normalArrow} = [thick, decoration={markings,mark=at position
   1 with {\arrow{Triangle[length=6pt, width=5.75pt]}}}, preaction = {decorate}, line width=1.25pt, shorten >= 5.5pt]

\begin{tikzpicture}[auto, line width=0.1em]
\node[block, left color=lightblue!50!white, right color=lightblue!10!white, at={(0,0)}] (coarse) {Coarse \\ Synchronization} ;
\node[block, left color=lightblue!50!white, right color=lightblue!10!white, below= 1.5em of coarse] (sro) {\acrshort{SRO} \\ Estimation};
\node[block, left color=orange!50!white, right color=orange!10!white, below= 1.5em of sro] (tdoa) {\acrshort{TDoA} \\ Estimation}  ;
\node[block, below= 1.5em of tdoa, left color=orange!50!white, right color=orange!10!white,] (distance) {Distance \\ Estimation} ;
\node[block, below = 1.5em of distance, left color=orange!50!white, right color=orange!10!white,] (doa) {\acrshort{DoA} \\ Estimation};
\node[block, right= 2.0em of doa, left color=red!50!white, right color=red!10!white,] (geo) {Geometry \\ Calibration};
\node[block, right= 2.0em of tdoa, left color=lightblue!50!white, right color=lightblue!10!white,] (sto) {\acrshort{STO} \\ Estimation};
\node[block, above= 1.5em of sto, left color=lightblue!50!white, right color=lightblue!10!white,] (sync){Synchronization};
\node[block, right=2.5em of sto, left color=red!50!white, right color=red!10!white,] (diary){Diarization};
\node[block, below=1.5em of diary, left color=lightgreen!50!white, right color=lightgreen!10!white,] (enh){Speech \\ Enhancement};
\node[block, below=1.5em of enh, left color=lightgreen!50!white, right color=lightgreen!10!white,] (asr){Speech \\ Recogniton};

\draw[vecArrow] ($(coarse.west)- (2.5em, 0)$) -- (coarse.west);
\draw[innerWhite] ($(coarse.west)- (2.5em, 0)$) -- (coarse.west);
\draw[vecArrow] (coarse.south) -- (sro);
\draw[innerWhite] (coarse.south) -- (sro);
\draw[normalArrow] (doa) -- (geo);
\draw[normalArrow] (tdoa) -- (sto);
\draw[normalArrow] (distance.east) -| (sto.south);
\draw[normalArrow] (distance.east) -| (geo.north);
\draw[vecArrow] ($(coarse.south) - (-.0725em, .3em)$) -| (sync.north);
\draw[innerWhite] ($(coarse.south) - (-.05em, .3em)$) -| (sync.north);
\draw[normalArrow] (sro) -- (sync);
\draw[normalArrow] (sto) -- (sync);
\draw[normalArrow] (diary) -- (enh);
\draw[vecArrow] (enh) -- (asr);
\draw[vecArrow] (sync.east) -| ($(diary.east) + (1.5em, 0)$) |- (enh.east);
\draw[innerWhite] (sync.east) -| ($(diary.east) + (1.5em, 0)$) |- (enh.east);
\draw[vecArrow] ($(diary.north) + (0, 2.8em)$) -- (diary.north);
\draw[innerWhite] ($(diary.north) + (0, 2.875em)$) -- (diary.north);
\draw[normalArrow] ($(sro.east) + (.75em, 0)$) |- ($(sro) !0.5! (tdoa)$) -| ($(sro.west) - (1.5em, 2.5em)$) |-  ($(tdoa.west) - (0, 0.5em)$);
\draw[normalArrow] ($(sro.east) + (.75em, 0)$) |- ($(sro) !0.5! (tdoa)$)-| ($(sro.west) - (1.5em, 2.5em)$) |-  ($(distance.west) - (0, 0.5em)$);
\draw[normalArrow] ($(sro.east) + (.75em, 0)$) |- ($(sro) !0.5! (tdoa)$)  -| ($(sro.west) - (1.5em, 2.5em)$) |-  ($(doa.west) - (0, 0.5em)$);
\draw[vecArrow] ($(coarse.south) - (0.04, .3em)$)  -| ($(coarse.south) - (6.5em, .4em)$) |-  ($(doa.west) + (0, 0.5em)$);
\draw[innerWhite] ($(coarse.south) - (0.03em, .3em)$)  -| ($(coarse.south) - (6.5em, .4em)$) |-  ($(doa.west) + (0, 0.5em)$);
\draw[vecArrow] ($(distance.west) + (-2.625em, 0.5em)$)--  ($(distance.west) + (0, 0.5em)$);
\draw[innerWhite] ($(distance.west) + (-2.7em, 0.5em)$) -- ($(distance.west) + (0, 0.5em)$);
\draw[vecArrow] ($(tdoa.west) + (-2.625em, 0.5em)$)--  ($(tdoa.west) + (0, 0.5em)$);
\draw[innerWhite] ($(tdoa.west) + (-2.7em, 0.5em)$) -- ($(tdoa.west) + (0, 0.5em)$);
\draw[normalArrow] (geo.east) -| ($(geo) !0.5! (diary)$) |- (diary.west);
\end{tikzpicture}

%% file: sections/source_sep.tex

\subsection{Source separation}
\label{subsec:source_sep}
Each speaker is now modeled by a component of a spatial mixture model. Additionally, a noise class is introduced that is assumed to be always active. The diarization estimates $\widehat{a}_i[\ell]$ are taken as time-varying class priors \cite{Ito2013permutation} of spatial mixture models \cite{Ito2016cACGMM}, one for each frequency bin with shared class priors, after normalization:
\begin{align}
    \pi_{i}[\ell] = \frac{\widehat{a}_{i}[\ell]}{\sum\limits_{\nu=1}^{I+1} \widehat{a}_{\nu}[\ell]}.
\end{align}
The mixture models are now initialized by taking these priors to be the initial frequency-independent class posterior probabilities $\gamma_{i}[\ell,k]$,  where $k \in \{1,\ldots , K\}$ is the frequency index.
The parameters of the mixture models are then optimized with the EM algorithm.
After convergence, the class posterior probability $\gamma_{i}[\ell,k]$ denotes the probability for 
source $i$ to be active at a given time-frame $\ell$ and frequency bin $k$.

Furthermore, the prior probabilities $\pi_{i}[\ell]$, after some temporal smoothing, are used to cut the meeting into segments and remove the noise class.
The posterior probabilities $\gamma_{i}[\ell, k]$ are used to extract the target speakers speech of a segment with a convolutional beamformer \cite{Boeddeker2020FactorizedConvBF, Nakatani2020ConvBF}. The posterior of the target speaker is used as the target mask, while the sum of all remaining class posteriors are used as distortion mask.

%% file: sections/experiments.tex
\section{Experiments}
\label{sec:exp}
We used a data set of $100$ simulated meeting scenarios to evaluate the proposed meeting transcription system.
For each scenario, a conference room, whose setup is visualized in~\cref{fig:setup}, was modeled according to the following scheme: 
First, the length $L_T$ and width $W_T$ of a rectangular conference table are drawn at random from uniform densities: $L_T {\sim} \mathcal{U}(\SI{1.5}{\metre},\SI{3.0}{\metre})$; $W_T {\sim} \mathcal{U}(\SI{1.5}{\metre},\SI{3.0}{\metre})$. 
Afterwards, $I{\sim} \mathcal{U}(3, 6)$ speaker positions are placed around the table so that a  distance between $\SI{0}{m}$ and $\SI{0.4}{m}$ to the edge of the table and a minimum distance of $\SI{0.5}{m}$ between the speakers is guaranteed.
The $J{=}3$ microphone arrays, consisting of $M{=}4$ microphones each and forming a square with $\SI{5}{\centi \metre}$ long edges, are placed on the table with random orientations and positions so that they are not colinear (avoidance of end-fire constellations) and have a minimum distance of $\SI{0.2}{m}$ from the edges of the table and from each other.
Finally, the table is randomly rotated and placed in a room, whose length and width are drawn from $\mathcal{U}(\SI{5}{\metre},\SI{7}{\metre})$ each, such that a minimum distance of $\SI{1}{m}$ to each wall is maintained.
All simulated rooms have a height of $\SI{3}{\metre}$ and a reverberation time randomly drawn from $\mathcal{U}(\SI{0.2}{\second},\SI{.5}{\second})$.
The microphone arrays and speakers are placed on a two-dimensional plane at a height of $\SI{1.6}{\metre}$.

For each conference room setup, a meeting of \SI{5}{min} duration was simulated.
Firstly, a set of speakers is randomly drawn from the \texttt{eval92} WSJ database~\cite{WSJ} and assigned to the speaker positions. 
Subsequently, a meeting is generated based on the set of speakers such that the speaking portions of all speakers are approximately equal.
Hereby, regions of a single speaker being active amount for $\SI{66}{\%}$ of the total duration, while two speakers are concurrently active for $\SI{21}{\%}$ of the time, and in the remaining time there is no speech activity. 
All recordings were reverberated via the image method~\cite{image_source} using the implementation of~\cite{habets_rir} and overlaid with an additive white sensor noise with an average \gls{SNR} drawn from  $\mathcal{U}(\SI{20}{dB}, \SI{30}{dB})$. 
The nominal sampling frequency of the meeting data is $f_s{=}\SI{16}{\kilo \hertz}$.
Finally, an \gls{STO}, which is drawn from $\mathcal{U}(\SI{0}{\second}, \SI{2}{\second})$, and a time-varying \gls{SRO}, whose average value is drawn from $\mathcal{U}(\SI{-100}{\acrshort{ppm}}, \SI{100}{\acrshort{ppm}})$, are generated (see \cite{Gburrek22} for more details).  
The \gls{SRO} is simulated using the \gls{STFT}-resampling method from \cite{Schmalenstroeer2018}.


\subsection{Baselines}
\label{subsec:baselines}
As two baselines we employ a single-channel system and a system solely using a single compact microphone array. 
The single-channel baseline system corresponds to the baseline system used in~\cite{Raj2021} consisting of a mask-based source separator followed by a diarization module.
The mask estimator in the separation model is a BLSTM with three layers, each with 600 units in each direction, followed by two fully connected layers.
The network is trained with the Graph-PIT \cite{vonNeumann2021_GraphPITGeneralizedPermutation} training scheme with the SA-tSDR \cite{vonNeumann2022_SaSdr} loss to produce two output streams that no longer contain speech overlaps.
During evaluation, the meeting data is cut  into temporally overlapping segments, and a  stitching approach \cite{20_Chen_LibriCSS} is used to concatenate the segments to the original meeting length.
After separation, an energy-based \gls{VAD} is used to extract single-speaker segments for diarization from both output streams. 
A 256-dimensional speaker embedding is extracted for each segment. 
Then, these embeddings are clustered with an agglomerative hierarchical clustering scheme to assign each segment to a speaker.
The embedding extractor is a ResNet34 trained on the VoxCeleb dataset \cite{19_Nagrani_VoxCeleb} as described in \cite{20_Tianyan_ResNet}.

The single-array baseline corresponds to a modified version of the proposed meeting transcription system.
Since a single array is used, the signal synchronization and geometry calibration systems are not needed, here.
Furthermore, the proposed diarization system is based on the distributed fashion of the \gls{WASN}.
Therefore, the spatial mixture model of the single-array baseline is initialized based on the estimated relative speaker positions w.r.t. the local coordinate system of the microphone array following from the \gls{DoA} and speaker-array distance estimates.
First, the relative source position estimates are clustered leading to a set of speaker position candidates. 
Afterwards, an energy-based \gls{VAD} is used to detect in which frames a speaker is active.
For all frames with speech activity the speaker at the speaker position candidate which is closest to the relative speaker position estimate of the frame is decided to be active.
Finally, the activity estimates are smoothed over time for each speaker position candidate. 

\subsection{Performance measures}
We use the \gls{DER} \cite{NIST2009_2009RT09Rich} to evaluate the performance of the proposed spatial diarization method.
Since the diarization system relies on the signal synchronization and the speaker localization performance, the \gls{DER} indirectly reflects the performance of the corresponding subsystems.
We use a collar of \SI{0.25}{\second} according to the specifications in~\cite{NIST2009_2009RT09Rich}.

In order to evaluate the source separation performance of our system we use the \gls{cpWER}~\cite{20_watanabe_chime6} as metric.
The \gls{ASR} results are obtained using the acoustic model from \cite{Drude_19_smswsj} that is openly available for a sampling rate of \SI{8}{\kilo\hertz}.
To be able to use it, the separated signals are downsampled after synchronization and diarization instead of retraining the model for \SI{16}{\kilo\hertz}.
We use an oracle diarization system to be able to compute the \gls{cpWER} for the single-channel baseline system.

\begin{table}[t!]
	\centering
	\caption{Diarization and transcription performance}
    \setlength{\tabcolsep}{0.5em}
	\begin{tabular}{ l c c S[round-precision=2,round-mode=places,table-format=2.2] S[round-precision=2,round-mode=places, table-format = 2.2] }
		\toprule	
		\textbf{System}  & \textbf{Sync.}  & \textbf{MM Init.}& \textbf{DER / \%}  &\textbf{WER / \%} \\
		\midrule
		Oracle Sep. & {---} & {---} & 0.00 & 8.598 \\
		\midrule
		Single-Ch. & {---} & {---} & 24.15 & 29.01 \\
		\midrule
		Single-Array  & {---} & Oracle & {---} & 16.336184188403036 \\  
		Single-Array  & {---} & Est. & 22.54 & 22.090950870388845 \\ 
		\midrule
		Multi-Array & Perfect & Oracle  & {---} & 13.916133298958672 \\  
		Multi-Array & Perfect & Dirichlet  & {---} & 15.758914059610146 \\ 
		Multi-Array  & Perfect & Est. & 7.35 & 14.194119691076412 \\ 
		\midrule
		Multi-Array & Coarse & Oracle & {---} & 19.63390985618688 \\  
		Multi-Array & Est. & Oracle  & {---} & 13.98675080985955 \\  
		\midrule
		Multi-Array & Est. & Dirichlet.  & {---} & 15.512313227892796 \\  
		Multi-Array  & Est. & Est. & 7.47 & 14.232230728705458 \\ 
		\bottomrule
	\end{tabular}
	\label{tab:results}
\end{table}

\subsection{Single-channel vs.~single-array vs.~multi-array}
\cref{tab:results} compares the proposed multi-array meeting transcription system with the single-channel and single-array baselines.
It can be seen that a better diarization as well as a better transcription can be achieved utilizing a set of distributed microphone arrays rather than a single compact microphone array.
In particular, the \gls{DER} can be significantly reduced.
This can be explained by the fact that the single-channel diarization suffers from errors in the single-channel source separation, while the single-array localization is unable to cope with overlap regions and is more unstable due to a lack of spatial diversity.

A comparison of the achieved \glspl{WER} reflects these limitation of the system utilizing a single microphone or a single microphone array, too.
It becomes apparent that the transcription performance of the proposed multi-array system, which uses signals synchronized based on the estimated \gls{SRO} and \gls{STO} (Est. Sync.) and the diarization results, is close to the performance of a multi-array system, which uses perfectly synchronous signals (Perfect Sync.) and the oracle activities of the speakers.
Furthermore, the need to compensate for an \gls{SRO} is shown by the fact that the performance of the multi-array system strongly deteriorates if the signals are only coarsely aligned based on the maximum of their cross-correlations over the complete meeting (Coarse Sync.).

\subsection{Informed mixture model initialization}
\cref{tab:results} further compares the performance of the proposed system to one that draws the initial values of the class posteriors of the mixture model at random from a Dirichlet distribution, while being otherwise identical to the proposed system.
It can be seen that an informed initialization with the estimated diarization improves the WER from 15.51\% to~14.23\%.


%% file: sections/conclusions.tex

\section{Conclusions and outlook}
\label{sec:conclusions}
We presented a system for meeting diarization and transcription for an ad-hoc acoustic sensor network consisting of multiple  initially unsynchronized microphone arrays. A key property of the system is that the synchronization and geometry estimation front-end delivers precise diarization information, which is used to ease the task of the subsequent enhancement stage. 
Simulations have shown that the proposed multi-array system is able to outperform a single-array system.
In future work we intend to remove the assumption of fixed speaker positions by combining spatial with spectral signatures of the speakers.